\documentclass[]{spie}  

\usepackage{amsmath,amsfonts,amssymb}
\usepackage{graphicx}
\usepackage[colorlinks=true, allcolors=blue]{hyperref}

\title{CCAT: Prime-Cam Control System and Related Software}

\author[a]{Alicia Middleton}
\author[b]{James Burgoyne}
\author[a]{Victoria Butler}
\author[b,c]{Scott Chapman}
\author[a]{Cody J. Duell}
\author[a]{Min Gao}
\author[d]{Graham Gibson}
\author[a]{Ben Keller}
\author[a]{Lawrence T. Lin}
\author[a]{Michael D. Niemack}
\author[a]{Darshan A. Patel}
\author[a]{Samantha Walker}
\author[a]{Yuhan Wang}
\author[b]{Ruixuan (Matt) Xie}
\affil[a]{Department of Physics, Cornell University, Ithaca, NY 14853, USA}
\affil[b]{Department  of Physics and Astronomy, University of British Columbia, Vancouver, BC, Canada}
\affil[c]{Department of Physics and Atmospheric Science, Dalhousie University, Halifax, NS, Canada}
\affil[d]{Department of Physics, University of Alberta, Edmonton, AB, Canada}

\authorinfo{Further author information: (Send correspondence to A.M.)\\A.M.: E-mail: amm542@cornell.edu}

\begin{document} 
\maketitle

\begin{abstract}
Prime-Cam is one of the primary instruments being developed for use by the CCAT collaboration with the Fred Young Submillimeter Telescope (FYST), which is undergoing construction at a height of 5600\,m in Cerro Chajnantor.  Prime-Cam can host up to seven instrument modules with kinetic inductance detectors (KIDs), including broadband modules ranging 280$\textendash$850\,GHz and imaging spectrometers for line-intensity mapping.  Prime-Cam requires a number of hardware devices for standard operation and data collection, and so a distributed control system must be developed for software integration of these devices and user-friendliness when controlling Prime-Cam in the field, especially during remote operation that will be typical during observations.  The Prime-Cam Control System (PCS) is being developed using the Observatory Control System (OCS) framework originally developed for Simons Observatory (SO) and is adapted for Prime-Cam and CCAT-specific requirements.  Within OCS/PCS, hardware devices are controlled individually by long-running software processes known as ``agents," which are typically deployed in Docker containers running in the background of one of the Prime-Cam network computers.  These agents are controlled by ``client" scripts through the command line or frontend tools like Nextline and PCS-web.  PCS also makes use of several external open source software tools for remote operations, data management, and logging.  This system is being tested in the lab in preparation for deployment in 2026 for use with Prime-Cam on FYST.
\end{abstract}


\section{INTRODUCTION}
\label{sec:intro}  
The Fred Young Submillimeter Telescope (FYST) is a 6-m aperture telescope located at a superior 5600\,m elevation on Cerro Chajnantor in the Chilean Atacama Desert\cite{CCAT-Prime_Collaboration_2023}.  Prime-Cam is a 1.8\,m cryostat receiver for FYST that is nearing deployment for use by the CCAT collaboration \cite{Vavagiakis_2018, Choi_2020}.  When fully populated, Prime-Cam can house up to seven instrument modules, including 280\cite{10.1117/12.2630115, lin2025ccatmodcamcryogenicperformance}, 350\cite{Keller2026}, 850\cite{10.1117/12.2630628}, and 410\cite{Patel2026} GHz broadband modules and Epoch of Reionization spectrometer (EoRspec) modules spanning 210-420 GHz for line-intensity mapping\cite{10.1117/12.3018477}.  In total, about 100,000 novel kinetic inductance detectors (KIDs) will be deployed in large arrays across all seven modules.  As of writing, first light is targeted for 2026 with the 280 and 350 GHz modules.

Prime-Cam requires an extensive network of computers, hardware devices, and software to support observatory operations, much of which will be performed by remote observers.  These operations include collection and monitoring of housekeeping (HK) data from various hardware devices, KID array readout with Radio Frequency System-on-Chip (RFSoC) FPGA boards, and full remote control of Prime-Cam network computers and devices.  Therefore, a distributed control system known as the Prime-Cam Control System (PCS) has been developed for the CCAT collaboration based on the Observatory Control System (OCS) framework originally developed for the Simons Observatory (SO).

   \begin{figure} [ht!]
   \begin{center}
   \begin{tabular}{c} 
   \includegraphics[width=\textwidth]{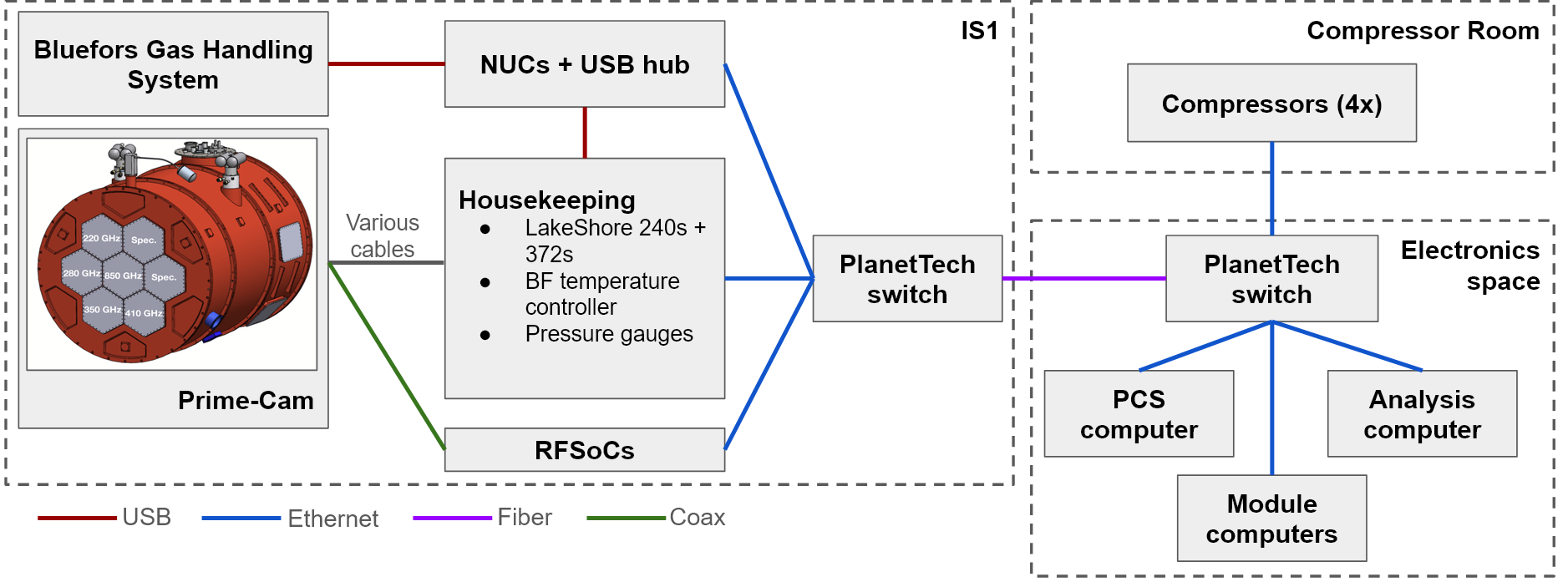}
   \end{tabular}
   \end{center}
   \caption[example] 
   { \label{fig:network} 
A diagram of the Prime-Cam computing network.  Prime-Cam and any hardware that must be directly connected to the instrument are located in IS1, while any computers and devices that can communicate over the network are in the climate-controlled electronics space, which is connected to IS1 via a fiber optics cable.}
   \end{figure} 

In this paper, we will first present an overview of the computing resources and hardware required for Prime-Cam operations in Section \ref{sec:computinghardware}.  Section \ref{sec:pcs} will provide an overview of PCS and associated software tools and interfaces designed for remote observers.  Then, in Section \ref{sec:datamanagement}, we will discuss the data management plan for ensuring that raw data from the Prime-Cam instrument is not lost and can be made into usable science data, before concluding in Section \ref{sec:conclusion}.

\section{Prime-Cam Computing Resources and Hardware}
\label{sec:computinghardware}

The Prime-Cam computing network consists of several main computer nodes as well as hardware for housekeeping and detector readout.  The main PCS computer, which will be expanded on in the following section, is a rack-mounted SuperMicro server known as \texttt{daq-primecam}.  This computer will run the majority of hardware agents that can connect to the network over Ethernet and will host PCS software tools and interfaces to be used by remote observers.  Other rack-mounted machines will include data collection servers; for initial deployment this includes \texttt{data-280GHz} and \texttt{data-350GHz}, which save and perform low-level processing on detector data from the 280 and 350 GHz modules, respectively.  All listed rack-mounted servers are currently running Ubuntu 22.04.  Other HK computers include two small form-factor Intel NUCs.  One NUC, \texttt{nuc-bluefors}, runs Windows 11 and handles the Bluefors software, which controls the gas handling system (GHS) for the dilution refrigerator (DR) and outputs the logs used by the Bluefors agent.  The other NUC, \texttt{nuc-daq}, runs Ubuntu 24.04 and handles USB connections for HK devices that use USB (as opposed to Ethernet) for data transfer.

The majority of the equipment and computing hardware for CCAT will be located in two compartments of FYST.  Prime-Cam itself as well as any hardware systems that must be connected to it directly will be in Instrument Space 1 (IS1). Any computers and hardware that can communicate over the network via Ethernet or fiber without needing to connect to Prime-Cam directly will be in the climate-controlled electronics space (ES).  We use a PlanetTech IGS-6325-20T4C4X network switch in each compartment to facilitate Ethernet connections between computers and hardware devices, and a 15\,m optical fiber cable will run between the switches in both spaces to facilitate high-speed data transfer while minimizing loss over a long distance.  A diagram of the Prime-Cam computing network is shown in Figure \ref{fig:network}.

   \begin{figure} [ht]
   \begin{center}
   \begin{tabular}{c} 
   \includegraphics[width=\textwidth]{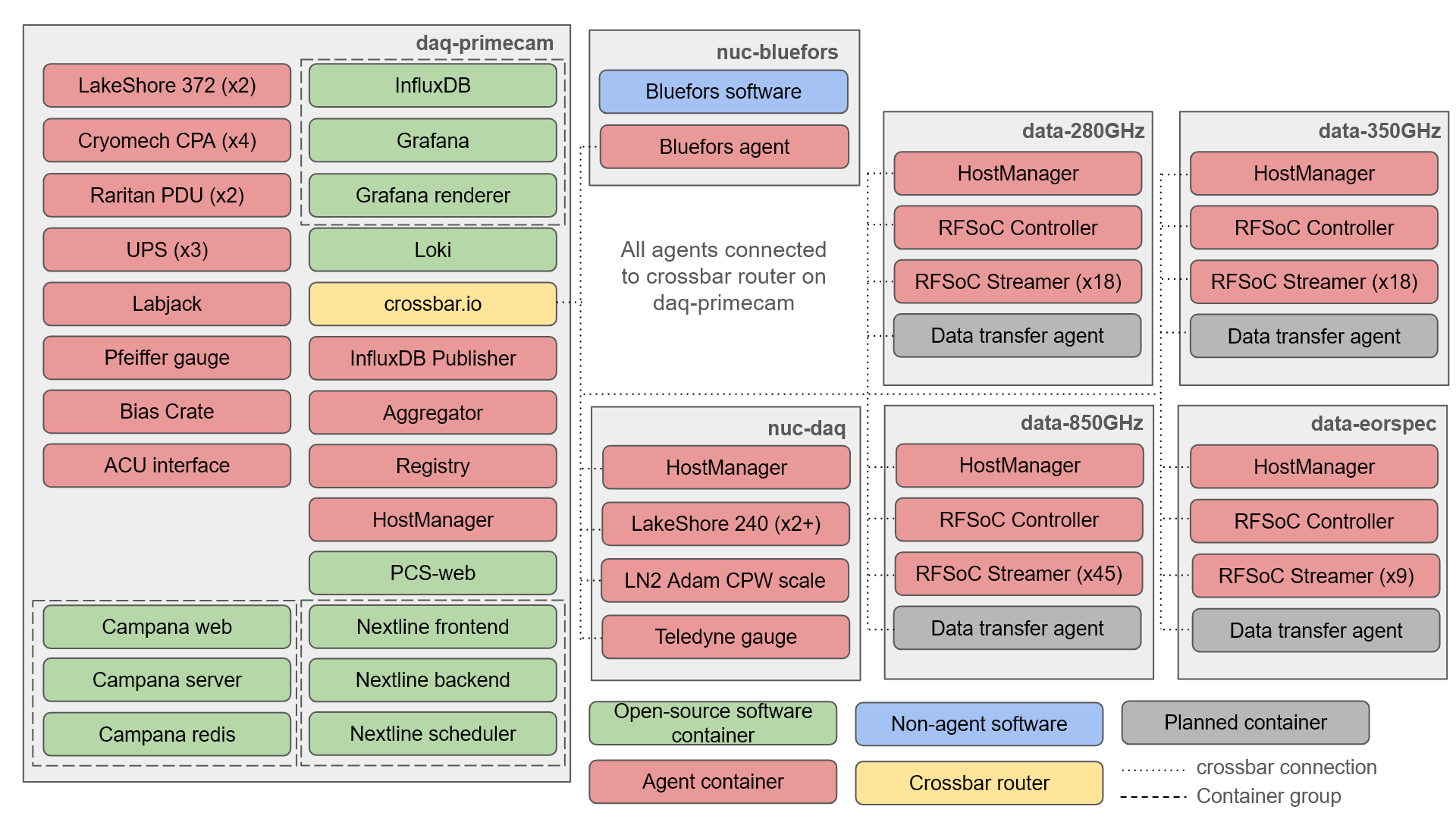}
   \end{tabular}
   \end{center}
   \caption[example] 
   { \label{fig:agents} 
Diagram of Prime-Cam network host computers and their associated PCS/OCS agents.  All agents are connected to the crossbar router on main PCS computer daq-primecam so they can be called by client scripts as well as through the PCS-web and Nextline browser interfaces.}
   \end{figure} 

\section{Prime-Cam Control System}
\label{sec:pcs}

   \begin{figure} [ht]
   \begin{center}
   \begin{tabular}{c} 
   \includegraphics[width=\textwidth]{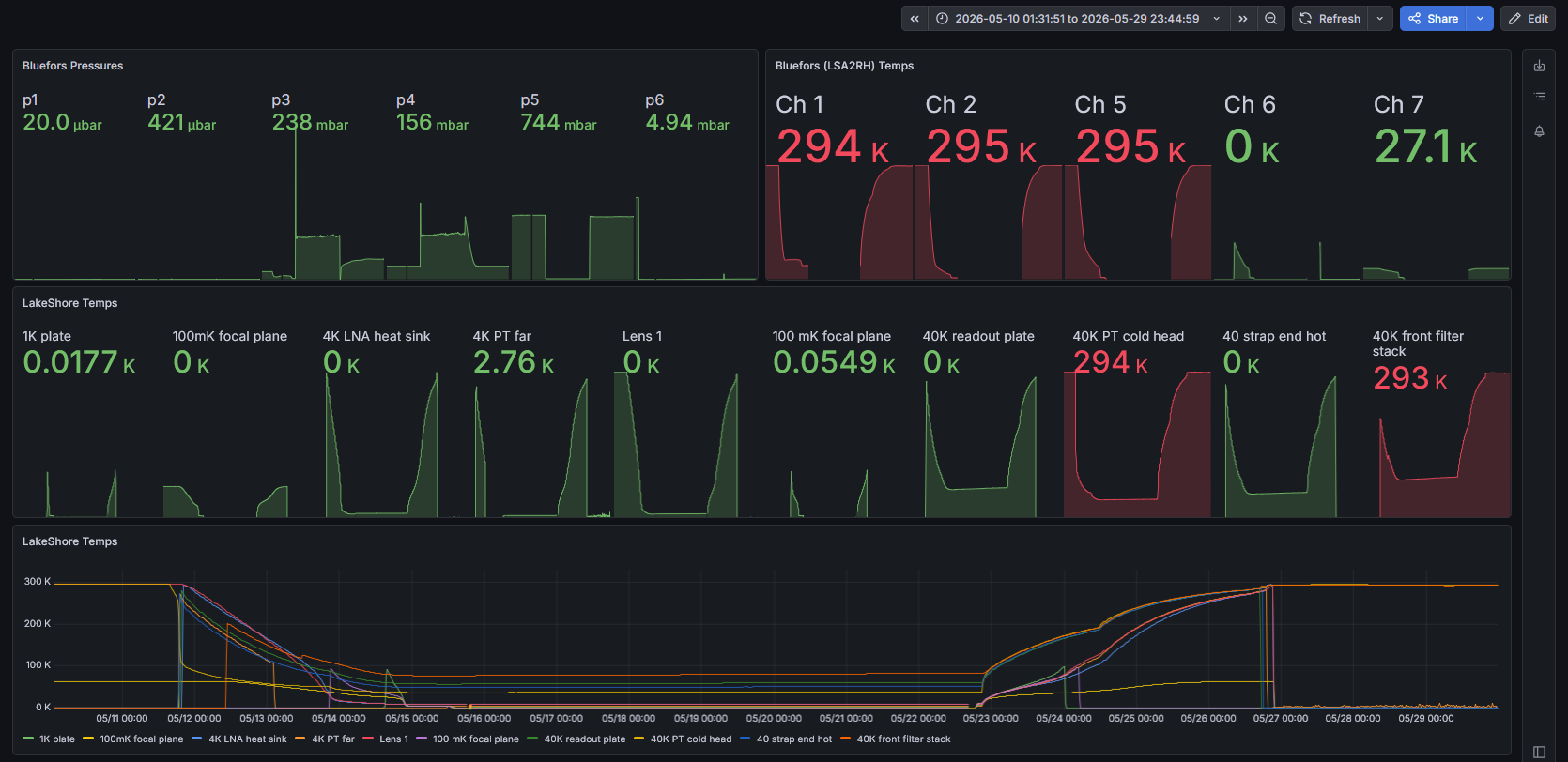}
   \end{tabular}
   \end{center}
   \caption[example] 
   { \label{fig:grafanadashboard} 
An example of part of a Grafana dashboard during a cooldown of prototype Mod-Cam system, similar to how it will be set up for Prime-Cam in the field. As shown, Grafana is used for monitoring cryogenic system temperatures and pressures and is used for configuring alerts to be sent to the group Slack and Campana alert system.}
   \end{figure} 

The Prime-Cam Control System (PCS) uses the Observatory Control System (OCS) framework originally developed for Simons Observatory (SO) \cite{10.1117/12.2561771} \cite{10.1117/12.3020710} and is adapted for Prime-Cam and CCAT-specific requirements. A full description of OCS as created for SO is found in Ref.~\citenum{10.1117/12.3020710}.  Within OCS, ``agents" refer to long-running software processes that control individual hardware and software devices.  These agents are controlled by ``client" scripts through the crossbar server, a WAMP router through which agents can be called.  Each agent has short-term tasks and long-term processes that can be called by clients or through frontend tools that will be described in the following subsections .  Agents are typically deployed in Docker containers\footnote{https://www.docker.com/} running in the background on a Prime-Cam network computer.  Agent scripts consist of short-term ``tasks" and long-term ``processes" which can be called by clients. For example, the LakeShore 240 agent, which controls the USB-connected LakeShore 240s, is running on \texttt{nuc-daq} and is connected to the PCS network through the crossbar router on \texttt{daq-primecam} as shown in Figure \ref{fig:agents}.

A common process for an PCS agent is data acquisition and logging. Data is written to OCS feeds, which are published through the crossbar server.  The InfluxDB publisher agent writes these feeds to an InfluxDB backend\footnote{https://www.influxdata.com/}, while the Aggregator agent records them in \texttt{.g3} files, a file format originally developed for SPT3G and in use with OCS on SO.  InfluxDB is designed to handle a large volume of time-ordered data and is a common backend for Grafana\footnote{https://grafana.com/}, an open-source web application for data visualization.  Docker plugin Loki also aggregates logs from the containers running PCS agents and allows for viewing in Grafana.  An example Grafana dashboard showing temperatures and pressures during a Mod-Cam cooldown is shown in Figure \ref{fig:grafanadashboard}.  Multiple dashboards have been created to monitor cryogenic housekeeping data, power usage, telescope pointing data, and detector status data.

The agents required for running an OCS instance, including the crossbar server, InfluxDB, and Grafana, are found in the SO \texttt{ocs} Github repository\footnote{https://github.com/simonsobs/ocs}.  Agents specifically developed for SO are tracked in the \texttt{socs} respository\footnote{https://github.com/simonsobs/socs}.  Several \texttt{socs} agents are used for interfacing with CCAT hardware devices and software, and \texttt{socs} agents that have been heavily edited and adapted for Prime-Cam are tracked in the CCAT \texttt{socs-ccat} repository\footnote{https://github.com/ccatobs/socs-ccat}.  CCAT-specific agents are found in the CCAT \texttt{pcs} respository\footnote{https://github.com/ccatobs/pcs}.  A diagram of all currently running and planned agents and software on the Prime-Cam network is shown in Figure \ref{fig:agents}.  

\subsection{Alarms and Alert Monitoring}
\label{subsec:alarms}

   \begin{figure} [ht]
   \begin{center}
   \begin{tabular}{c} 
   \includegraphics[width=\textwidth]{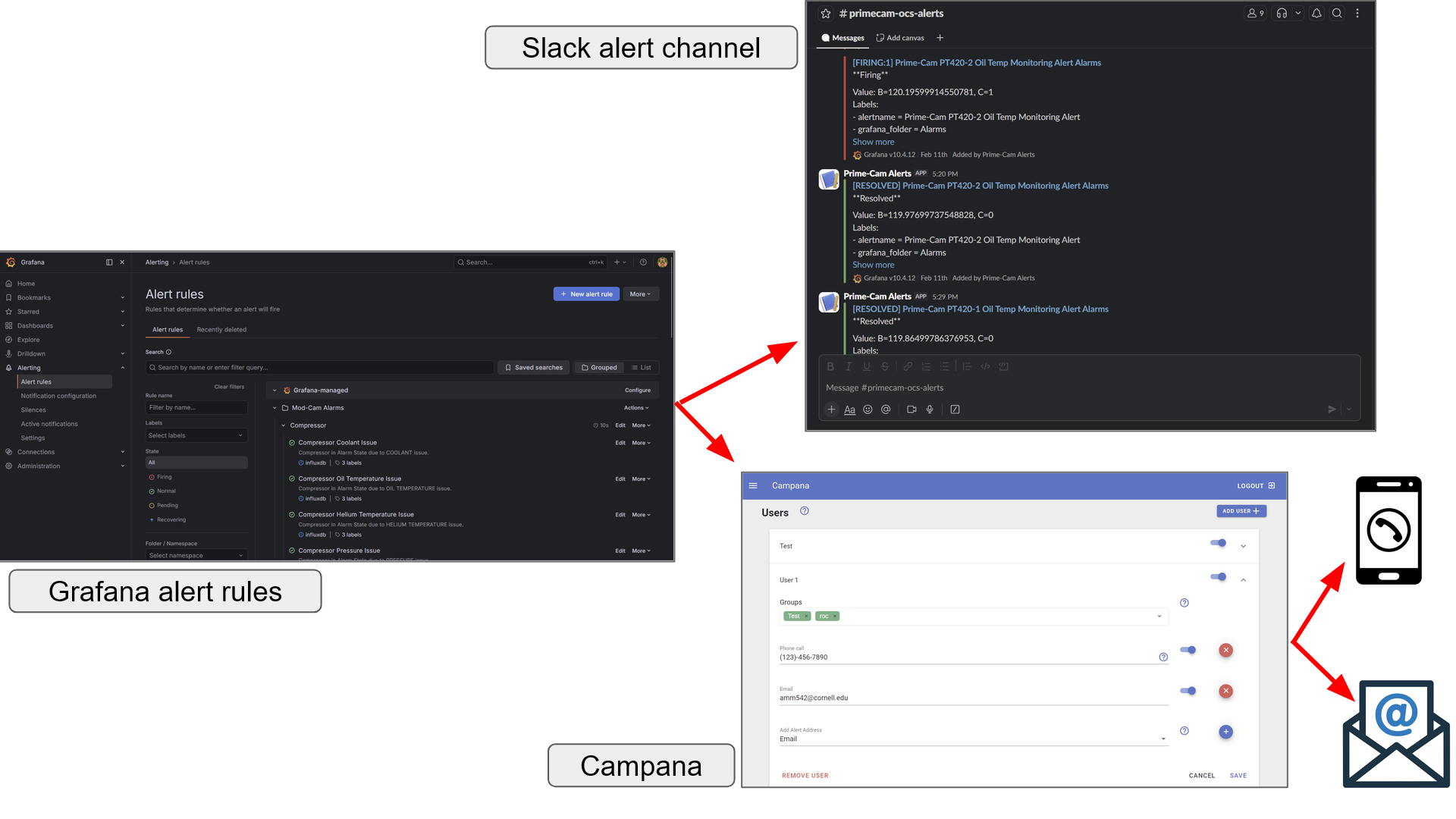}
   \end{tabular}
   \end{center}
   \caption[example] 
   { \label{fig:alerting} 
The alerting process for Prime-Cam.  First, alerts are configured in the Grafana alert rules dashboard.  Then, using webhooks for a Slack application, Grafana sends an alert notification to a Prime-Cam alerting channel in the CCAT collaboration Slack.  If the alarm is time-sensitive or urgent, a notification is also sent to Campana, which has the ability to email, SMS message, or call specific users or groups of users so the issue can tended to quickly.}
   \end{figure} 

A complex system like Prime-Cam can run into various problems that if left attended, can lead to interruptions in operations or even long-term damage to the instrument. For example, if pressure or temperature is very high in some part of the cryostat, or if data acquisition and logging is interrupted due to problems such as an agent going down or computer memory filling up, we will want the correct personnel to be notified. Therefore, a robust alerting system must be implemented so that any issue can be tended to as quickly as possible.  The alerting system for Prime-Cam is modeled after the alerting system for SO as described in Ref.~\citenum{10.1117/12.3018481}.

Grafana alerts are configured in the ``Alert rules" tab in Grafana, where conditions are selected from data feeds for alerting for any PCS agent.  These alerts are sent to the proper channels via configuration in the ``Notification configuration" tab.  Grafana monitors these alert rules for whether the conditions are normal or if the alert should be triggered.  Currently, firing alerts are sent to a group Slack channel. This Slack notification system is fully integrated with PCS and has been tested extensively with both Prime-Cam and the testbed Mod-Cam instrument in the lab. Work is currently being done on implementing Campana, SO's more involved alarm system which can send emails, SMS messages, and phone calls to the proper channels, allowing for quicker response in time-sensitive situations.  A schematic showing the Prime-Cam alerting process is shown in Figure \ref{fig:alerting}.

\subsection{PCS-web}
\label{pcsweb}

   \begin{figure} [p]
   \begin{center}
   \begin{tabular}{c} 
   \includegraphics[width=\textwidth]{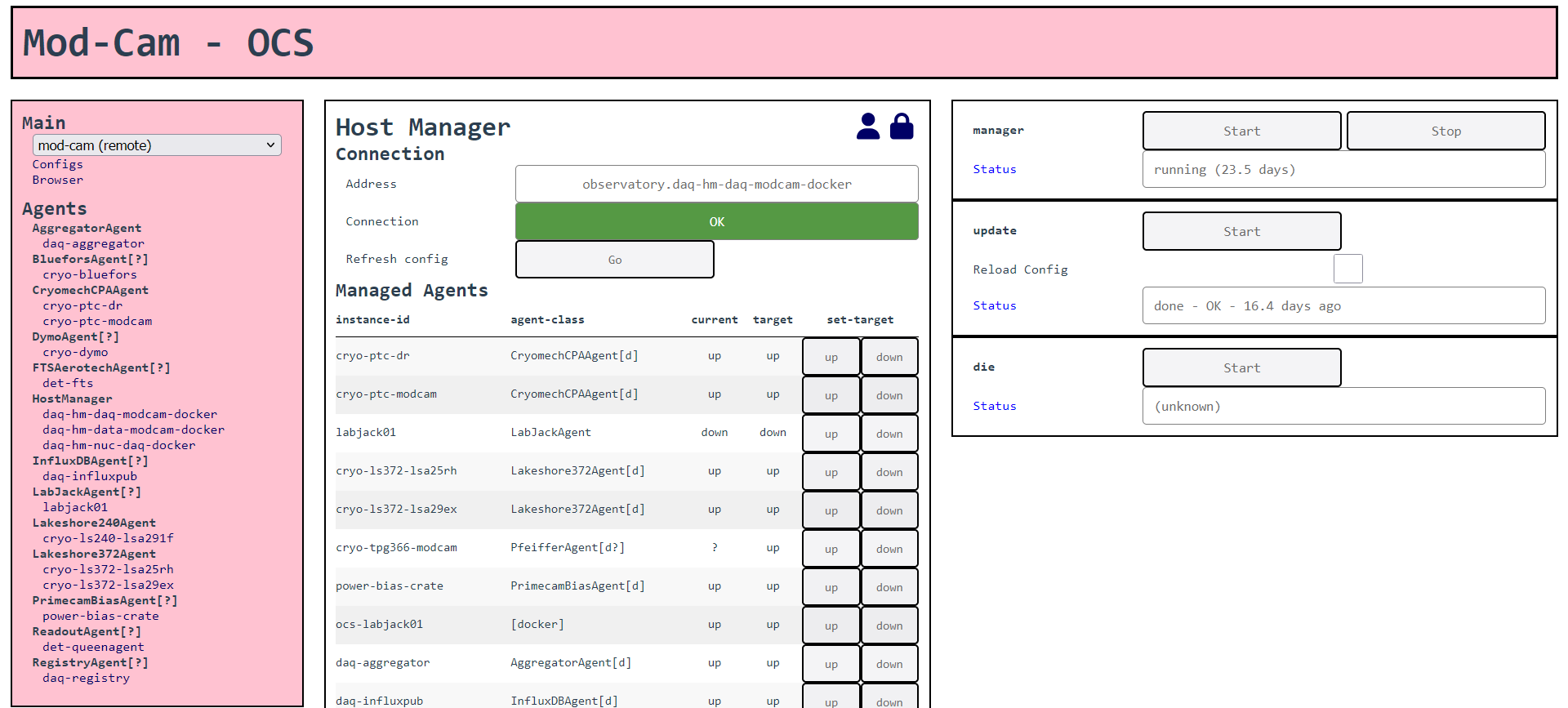}
   \end{tabular}
   \begin{tabular}{c} 
   \includegraphics[width=\textwidth]{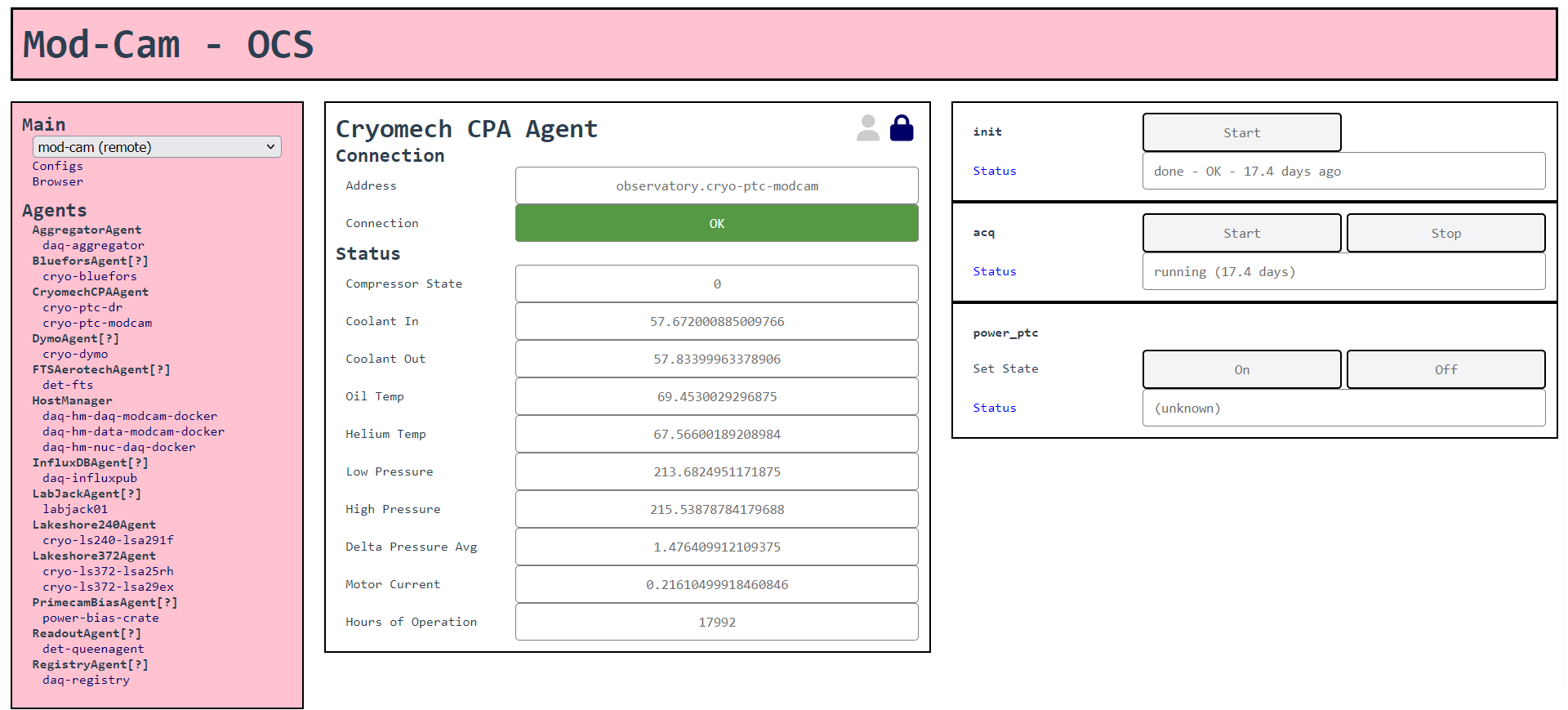}
   \end{tabular}
   \end{center}
   \caption[example] 
   { \label{fig:pcsweb} 
Top: The HostManager panel in PCS-web.  This panel lists all agents running on a machine and allows the user to start/stop specific agents.  Bottom:  The Cryomech CPA agent panel in PCS-web.  This panel uses functions of the Cryomech compressor agent to allow the user to have remote control of the compressors, including ability to turn the compressor on and off.  It also reads in and shows a snapshot of important values like the state of compressor (on or off), coolant in and out temperature, oil temperature, and helium temperature.}
   \end{figure} 

Although it is possible to control PCS agents using client scripts and the command line, a more user-friendly solution makes remote operations simpler for observers unfamiliar with PCS scripts and Docker commands. PCS-web is a web interface that allows remote control of PCS agents through buttons and text inputs, eliminating the need to be familiar with the aforementioned systems.

PCS-web uses the OCS-web interface developed for SO\footnote{https://github.com/simonsobs/ocs-web} as a base and has been adapted for CCAT with Prime-Cam-specific panels and features.  It communicates with agents through a configurable crossbar router, and each agent has its own panel that can be selected from a sidebar menu.  An individual agent panel includes a snapshot of the agent's status as well as a list of the agent's tasks and processes that can be executed in the interface, as shown in the example PCS-web agent panels in Figure \ref{fig:pcsweb}. Updates to PCS-web are tracked in the \texttt{pcs-web} repository for CCAT\footnote{https://github.com/ccatobs/pcs-web}.  The PCS-web interface has been extensively integrated and tested in the lab with both Prime-Cam and the testbed Mod-Cam instrument.

\subsection{Nextline}
\label{subsec:nextline}

   \begin{figure} [ht]
   \begin{center}
   \begin{tabular}{c} 
   \includegraphics[width=\textwidth]{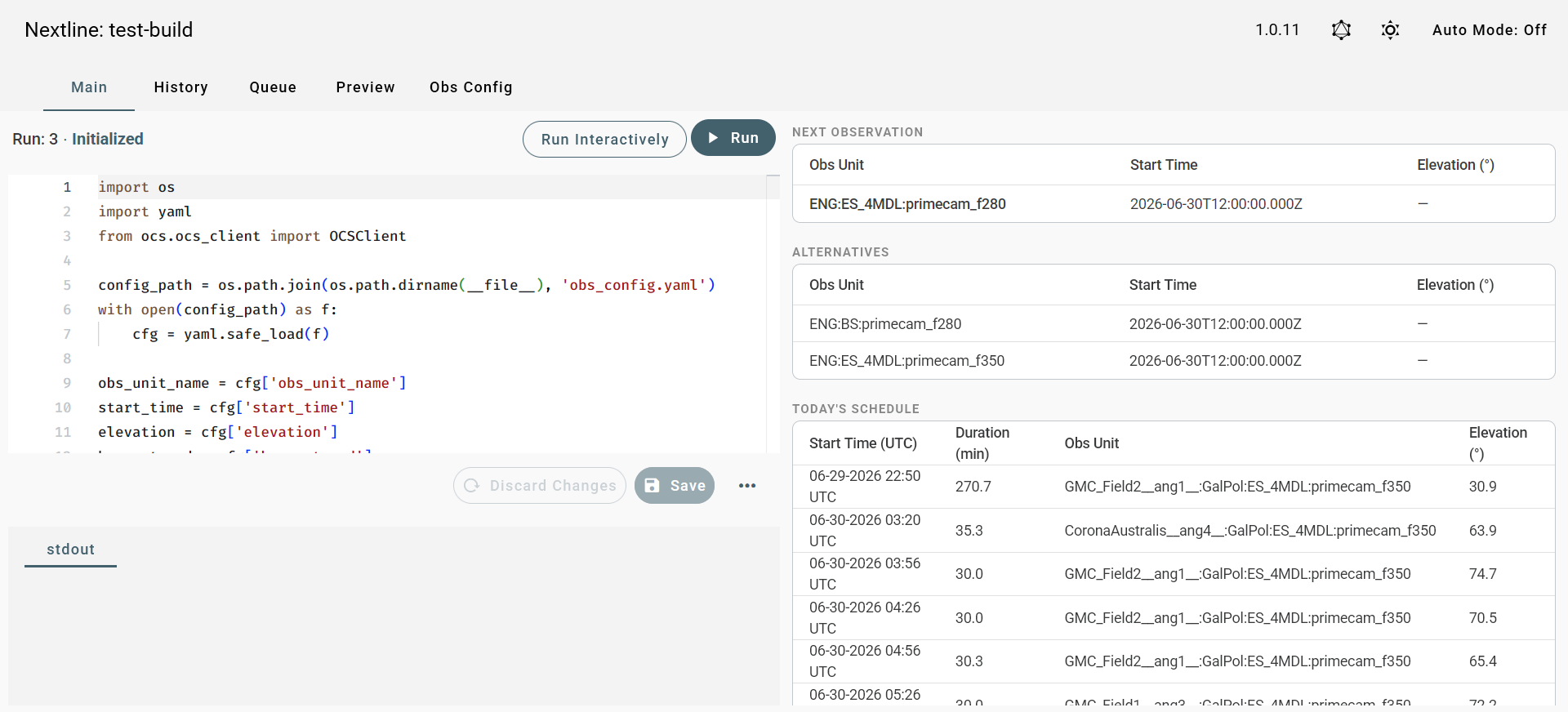}
   \end{tabular}
   \end{center}
   \caption[example] 
   { \label{fig:nextline} 
The Nextline browser interface for running generated observing scripts line-by-line.  The left side shows an example observing script, while the right shows the next observation, alternative observations, and the 24-hour schedule.}
   \end{figure} 

\setcounter{footnote}{0}

Nextline is another remote observing tool borrowed from SO that is complementary to the PCS-web interface. Nextline is a web interface with the ability to generate observing scripts from the CCAT Operations Database (OpsDB) and observatory scheduler and run them line-by-line by interfacing with PCS agents.  Observing script generation for CCAT using Nextline will be further discussed in Section \ref{sec:datamanagement}. While a remote observer can use PCS-web to run specific tasks and processes for individual agents, Nextline allows for more complex client scripts involving multiple agents and several sequential tasks and processes.

Both the frontend and the backend of SO's Nextline interface are being adapted for CCAT due to the different requirements of the SO versus the CCAT schedulers. Updates to the Nextline frontend browser interface for CCAT are tracked in the \texttt{nextline-web-ccat} repository\footnote{https://github.com/ccatobs/nextline-web-ccat}, which is forked from SO's \texttt{nextline-web} \footnote{https://github.com/simonsobs/nextline-web}.  The backend consists of two parts: the \texttt{nextline-graphql} backend API server from SO\footnote{https://github.com/nextline-dev/nextline-graphql}, and the CCAT scheduler server\footnote{https://github.com/ccatobs/ccat-scheduler-server}.

\subsection{Prime-Cam Remote Services Dashboard}

\begin{figure} [ht!]
   \begin{center}
   \begin{tabular}{c} 
   \includegraphics[width=\textwidth]{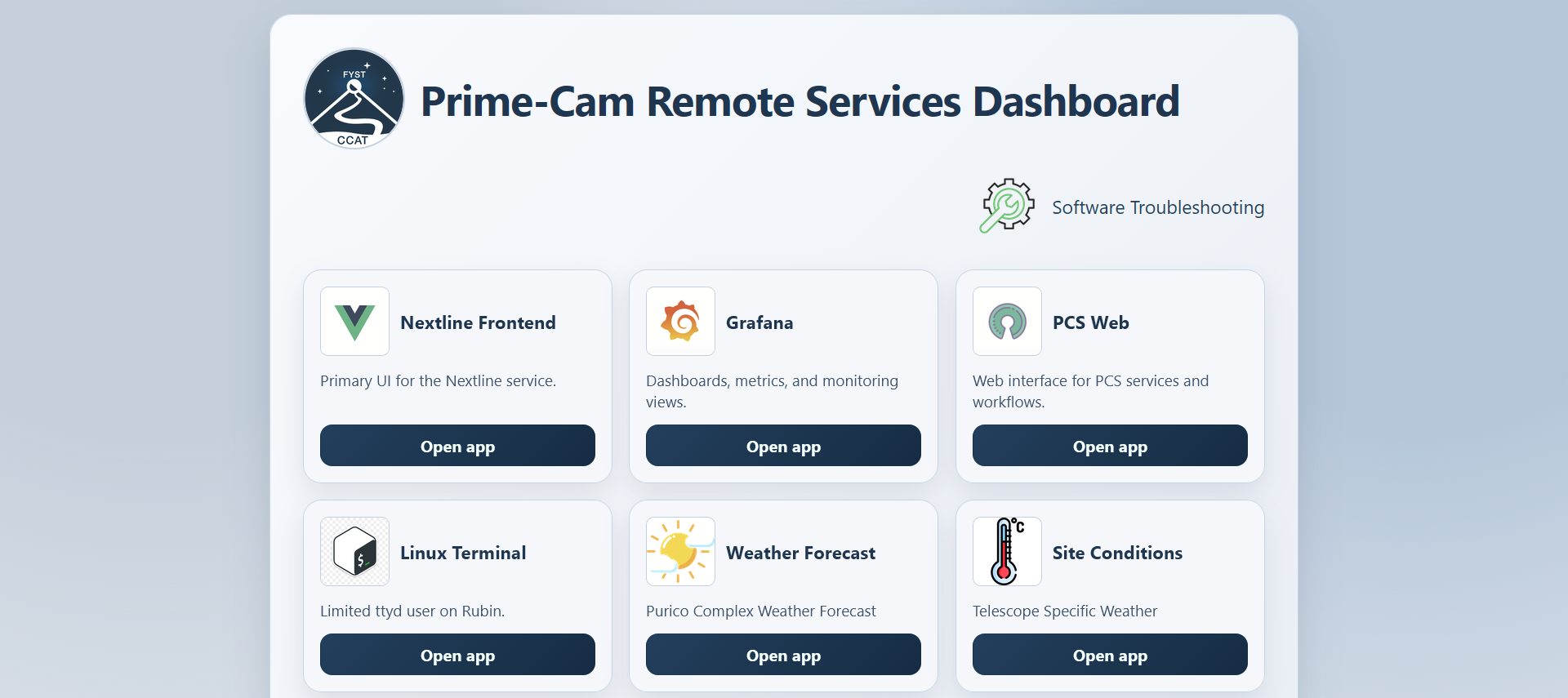}
   \end{tabular}
   \end{center}
   \caption[example] 
   { \label{fig:remotedashboard} 
The CCAT remote services dashboard.  This web-accessible landing page eliminates the need for remote users to tunnel in to individual interfaces.}
   \end{figure}

For ease of use by remote observers, we have created a remote services dashboard as a landing page for accessing the aforementioned browser interface tools, as shown in Figure \ref{fig:remotedashboard}.  This dashboard eliminates the need for tunneling in to individual interfaces, which quickly becomes complicated for users unfamiliar with the command line.

\section{Data Management}
\label{sec:datamanagement}

Due to the large volume of time-ordered data from numerous sources, a robust data management plan must be in place.  The three major data sources are HK data, telescope pointing data, and detector data. 

HK data acquired by PCS agents is constantly logged and updated in Grafana for live observer viewing as mentioned in Section \ref{sec:pcs}, and is stored in timestamped \texttt{.g3} files on the main PCS computer via the Aggregator agent. 

Significant telescope data is generated in the Nextline interface from the observing scripts.  Observing scripts generated in the Nextline interface combine input from the OpsDB and observatory scheduler.  OpsDB tracks possible observation targets and stores them as ObsUnit objects that contain  information about the field and scan pattern.  The scheduler itself generates the next best observing target as well as a list of five alternatives for a 24-hour period.  By referencing to OpsDB, a PCS client script (see Figure \ref{fig:nextline} left panel) is generated from the target ObsUnits (Figure \ref{fig:nextline} right table) with commands for interfacing with the ACU interface agent for telescope control and the RFSoC controller agent for detector readout.

The ACU interface agent takes in commanded coordinates to pass a trajectory along to the telescope control system (TCS) by communicating with the CCAT observatory control system.  More information about CCAT-specific scan patterns can be found in the \texttt{fyst-trajectories}\footnote{https://github.com/ccatobs/fyst-trajectories} repository.  The agent records azimuth and elevation pointing information which is generated at 200 Hz as UDP packets.  This data is likewise stored on the main PCS computer as timestamped \texttt{.g3} files with the HK data, and can be monitored and logged in Grafana.

The largest data volume in our instrument is handled by the RFSoC controller agents, which mediate detector readout and run on the corresponding packet-capture computers.  After VNA and target sweeps are acquired, time-ordered packets are streamed at a rate of about 488 Hz.  Using the RFSoC streamer agent, these packets are captured by the packet-capture computer for the corresponding module and the data is saved as \texttt{.g3} files on those computers.  A more detailed discussion of controlling the CCAT KID arrays with RFSoCs is found in Ref.~\citenum{patel2025ccatreadout10000280}.

The timestamped \texttt{.g3} files containing HK, telescope, and detector data are then combined and stored on an on-site computer, where quality checks will then be performed and a quick-look mapmaker will ensure data quality.  The data is then transferred to the data center in Cologne where it will be uploaded to RAMSES, a high-powered computing cluster operated by University of Cologne, for bookbinding and mapmaking.  When the data is confirmed to be stored off-site, it can then be deleted from on-site computers to free up storage for future observations.

\begin{figure} [ht]
   \begin{center}
   \begin{tabular}{c} 
   \includegraphics[width=\textwidth]{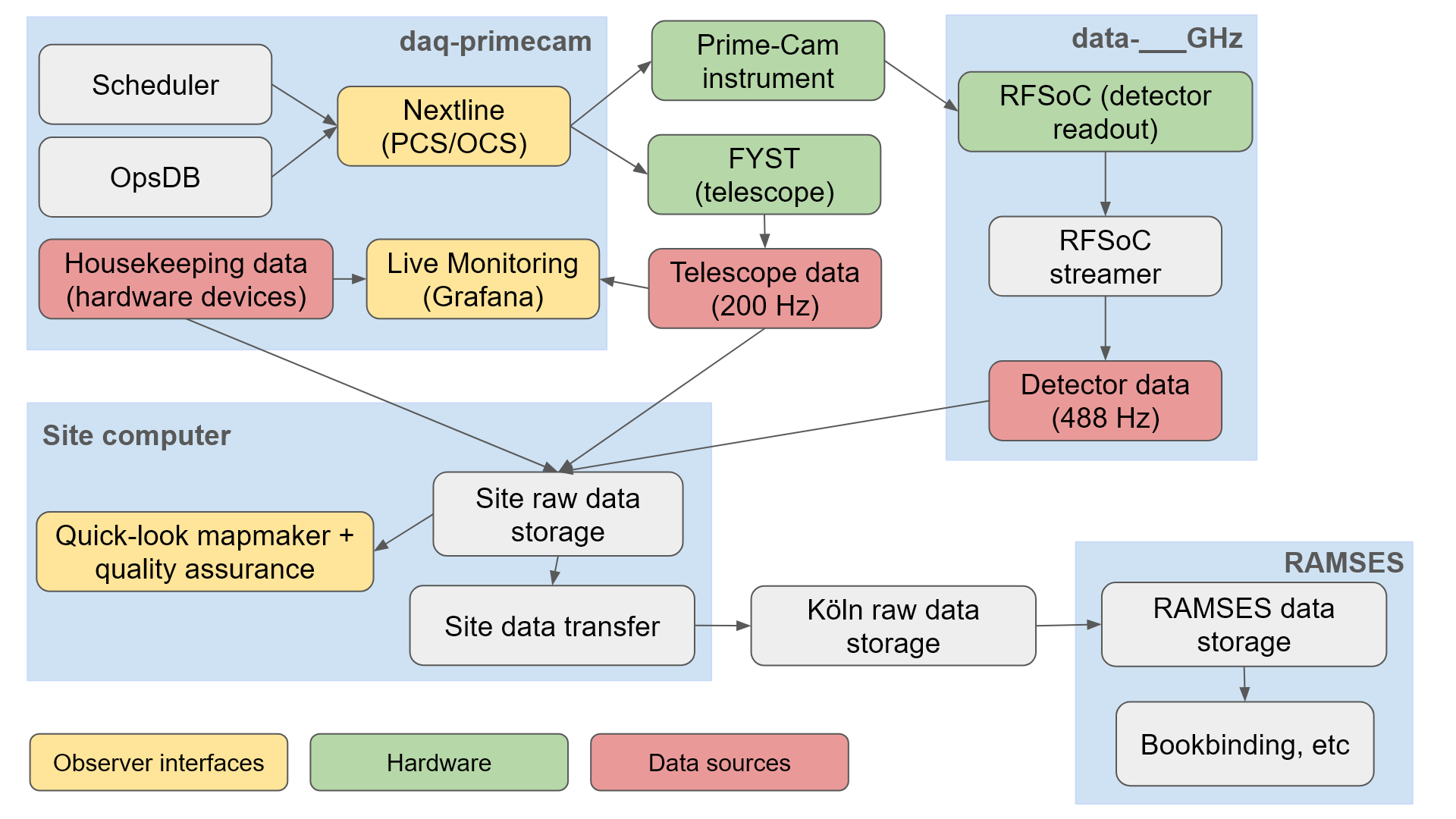}
   \end{tabular}
   \end{center}
   \caption[example] 
   { \label{fig:datamangement} 
A schematic showing a simplified version of the CCAT data management plan. A large volume of time-ordered housekeeping, telescope, and detector data from numerous sources must be kept track of during Prime-Cam operations to ensure data is not lost and can become usable science data.  Each blue box represents a Prime-Cam network computer, including main PCS computer \texttt{daq-primecam} and module data collection and RFSoC control computers as represented by the \texttt{data-\_\_\_GHz} naming convention for these servers.}
   \end{figure}

A simplified schematic of data management plan for CCAT is shown in Figure \ref{fig:datamangement}. This evolving data management process is currently undergoing lab testing and will adapt as needed as Prime-Cam is deployed with FYST and real data is acquired from the observatory.

\section{Conclusion}
\label{sec:conclusion}
In this paper, we presented an overview of the Prime-Cam Control System which is under development for use by the CCAT collaboration as we rapidly approach first light.  PCS builds on the well-tested and integrated OCS framework developed for SO by taking advantage of its distributed modular design to seamlessly adapt to CCAT and Prime-Cam-specific requirements.  A variety of software tools within PCS are designed to facilitate remote control, live monitoring, and data management for observations with Prime-Cam. User-friendly interfaces such as PCS-web, Nextline, and the CCAT Remote Services dashboard minimize the required training for remote observers.

This system has been extensively tested in the lab with both Prime-Cam and Mod-Cam. New hardware and software components are continuously integrated as the project evolves, with updates tracked in various CCAT Github repositories so that features are easily implemented across the Prime-Cam network.  PCS will continue to evolve as Prime-Cam is deployed.

\acknowledgments 
  
The CCAT project, FYST and Prime-Cam instrument have been supported by generous contributions from the Fred M. Young, Jr. Charitable Trust, Cornell University, Duke University, and the Canada Foundation for Innovation and the Provinces of Ontario, Alberta, and British Columbia. The construction of the FYST telescope was supported by the Gro{\ss}ger{\"a}te-Programm of the German Science Foundation (Deutsche Forschungsgemeinschaft, DFG) under grant INST 216/733-1 FUGG, as well as funding from Universit{\"a}t zu K{\"o}ln, Universit{\"a}t Bonn, and the Max Planck Institut f{\"u}r Astrophysik, Garching.

The construction of EoR-Spec is supported by NSF grant AST-2009767. The construction of the 350 GHz instrument module for Prime-Cam was supported by NSF grant AST-2117631. The construction of the 850 GHz instrument module for Prime-Cam is supported by CFI grants: 39656 and 46097 and Canadian provincial matching funds. The completion and deployment of the Prime-Cam instrument with the initial instrument modules is supported by a generous contribution from Alex Gerko, Founder and CEO of XTX Markets.

S. Walker acknowledges support by the National Science Foundation Award No. 2503181.
\bibliography{report} 
\bibliographystyle{spiebib} 

\end{document}